\documentclass[showpacs,amsmath,amssymb]{revtex4}

\usepackage{graphics}
\usepackage[dvips]{graphicx}

\def\sqr#1#2{{\vcenter{\hrule height.#2pt\hbox{\vrule width.#2pt height#1pt
\kern#1pt \vrule width.#2pt}\hrule height.#2pt}}}
\def\square{\mathchoice\sqr64\sqr64\sqr{4.2}3\sqr{3.0}3}

\begin{document}

\title{Cosmological simulations of the Santa Barbara cluster: the influence of scalar fields}

\author{M.A. Rodr\'\i guez-Meza}

\address{Depto. de F\'{\i}sica, Instituto Nacional de Investigaciones
Nucleares, Col. Escand\'on, Apdo. Postal 18-1027, 11801 M\'{e}xico D.F.
 \\
mar@nuclear.inin.mx; http://astro.inin.mx/mar
}

\date{\today}



\begin{abstract}
We present numerical $N$-body simulation studies of large-scale structure formation.
The main purpose of these studies is to analyze the several models of dark matter
and the role they played in the process of large-scale structure formation. 
We analyze in this work a flat cold dark matter dominated model known as the 
{\em Santa Barbara cluster}.
We compare the results for this model using the standard Newtonian limit of general relativity
with the corresponding results of using the Newtonian limit of scalar-tensor theories.
An specific model
is the one that considers that the scalar field is non-minimally coupled to the Ricci
scalar in the Einstein-Hilbert Lagrangian. Comparisons of the models are done
showing results of rotation curves, density profiles, and velocity dispersions for halos
formed at z=0. 
We analyze, in particular, the Santa Barbara cluster and its possible equation of
state.
\end{abstract}

\maketitle


\section{Introduction}

The Santa Barbara (SB) cluster model was introduced by Frenk et al. \cite{Frenk1999}
in order to study in a systematic way a flat cold dark matter (FCDM) dominated universe
using a variety of numerical codes. 
The main goal of this comparison was to asses the
reliability of cosmological simulations of clusters in one of the simplest astrophysical
relevant case. They compared the images and global properties of the cluster obtained
by the different numerical codes. Heitmann et al. (2005) \cite{Heitmann2005} analyze 
again the cluster with others new numerical codes and with a similar purpose 
and now it has become one of the standard cases of study. 
Given that this is a standard case to test we have decided to analyze it in the framework
of the scalar-tensor theories (SST) of gravity.

Scalar fields have been considered as one of the best possible ways to modify
gravity. The work by Nordstr\"om, published before general relativity, formulated a
conformally flat scalar theory of gravity \cite{Nordstrom1912}, and finally,
the scalar field role in gravity has been stablished 
since the
pioneering work of Jordan, Brans, and Dicke\cite{Jordan, BransDicke}.
Nowadays they are considered as a mechanism for inflation\cite{Copeland2004};
the dark matter component of galaxies\cite{Guzman2000};
the quientessence field to explain dark energy in the universe\cite{Axel2004}.
The main goal of this work is to study the large scale structure formation 
where  the usual approach is that the evolution of 
the initial primordial fluctuation energy density fields evolve following 
Newtonian mechanics in an expanding background\cite{Peebles1980}.
The force between particles are the standard Newtonian gravitational force.
Now, we will see that we can introduce the scalar fields by adding a term in this force.
This force will turn out to be of Yukawa type with two parameters 
($\alpha$, $\lambda$)\cite{mar2004}.
For so many years this kind of force, the so called fifth force, 
was thoroughly studied theoretically\cite{Pimentel} 
and many experiments were done to constrain the Yukawa 
parameters\cite{Sudarsky}.
We have also been studying, in the past years, the effects of 
this kind of force on some astrophysical 
phenomena\cite{mar2004,Rodriguez2001,mar2005,jorge2007} and in cosmological
simulations\cite{mar2007,mar2008}.
The Yukawa force comes as a Newtonian limit of a scalar-tensor theory with the scalar 
field non-minimally coupled to gravitation\cite{Helbig} although other alternatives can be 
found\cite{Nusser2005}.

Our general purpose is to find the role these scalar fields play on the large scale structure 
formation processes. In particular,
in this work we present some results about the role scalar fields play 
on cosmological simulations that form the SB cluster.

We start by discussing a FCDM model and the general approach in $N$-body
simulations (See Bertschinger\cite{Bertschinger1998} for details). 
Then, we present the modifications
we need to do to consider the effects of a static scalar field and we show the
results of this theory for the FCDM model that form the Santa Barabara cluster
\cite{Frenk1999,Heitmann2005}. 
To perform the simulations we have modified a standard 
treecode the author has developed \cite{Gabbasov2006} and the Gadget 1 \cite{Springel2001} 
(see also \url{http://www.astro.inin.mx/mar})
in order to take into account the contribution of the Yukawa potential. We finish this paper
by discussing how we can obtain the equation of state for a dark matter halo in the framework
of general relativity and its Newtonian limit.

\section{Evolution equations for a CMD universe}
\subsection{General Scalar-tensor theory}
The Einstein equations for a typical scalar--tensor theory with a massive scalar
field non-minimally coupled to the geometry are given by
\begin{eqnarray}
R_{\mu\nu} - \frac{1}{2} g_{\mu\nu} R &=& \frac{1}{\phi}
\left[ 8 \pi T_{\mu\nu} + \frac{1}{2} V g_{\mu\nu}
+ \frac{\omega}{\phi} \partial_\mu \phi \partial_\nu \phi
\right. \nonumber \\
&& \left. -\frac{1}{2} \frac{\omega}{\phi}(\partial \phi)^2 g_{\mu\nu}
+ \phi_{;\mu\nu} - g_{\mu\nu} \, \square \phi \frac{\mbox{}}{\mbox{}}
\right] \; , 
\end{eqnarray}
and the scalar field equation
\begin{equation}
\square \phi + \frac{\phi V' - 2V}{3+2\omega} = \frac{1}{3+2\omega} \left[
	8\pi T -\omega' (\partial \phi)^2 \right] \, ,
\end{equation}
where $()' \equiv \frac{\partial }{\partial \phi}$. 
Here $g_{\mu\nu}$ is the metric, $R$ is the Ricci's scalar, $R_{\mu\nu}$ the Ricci's tensor,
$T_{\mu\nu}$ is the energy-momentum tensor,
and $\omega(\phi)$ and
$V(\phi)$ are arbitrary functions of the scalar field $\phi$. We will not consider a cosmological
constant contribution in this work.

\subsection{Newtonian approximation of STT}
The study of large-scale formation in the universe is greatly simplified by the fact that a
limiting approximation of general relativity, Newtonian mechanics, applies in a region
small compared to the Hubble length $cH^{-1}$ ($cH_0^{-1}\approx 3000 h^{-1}$ Mpc, where 
$c$ is the speed of light, $H_0=100 h$ km/s/Mpc, 
is Hubble's constant at this epoch and $h\approx 0.7$), and large compared to the Schwarzschild
radii of any collapsed objects. The rest of the universe affect the region only through a tidal field.
The length scale $cH_0^{-1}$ is of the order of the largest scales currently accessible in 
cosmological observations and $H_0^{-1} \approx 10^{10}h^{-1}$ yr 
characterizes the evolutionary time scale of the universe.

Therefore we need to describe
the STT theory in its Newtonian approximation, that is, where gravity 
and the scalar fields are weak (and time independent) and velocities of dark matter 
particles are
non-relativistic.  We expect to have small deviations of the metric with respect to
Minkowski metric and of
the scalar field around the background field, defined here as
$\langle \phi \rangle$ and can be understood as the scalar field beyond all matter.
If one defines the perturbations $\bar{\phi} = \phi - \langle \phi \rangle$ and
$ h_{\mu\nu} = g_{\mu\nu} - \eta_{\mu\nu}$,
where $\eta_{\mu\nu}$ is the Minkowski metric, the Newtonian approximation
gives \cite{Helbig}
\begin{eqnarray}
R_{00} = \frac{1}{2} \nabla^2 h_{00} &=& \frac{G_N}{1+\alpha} 4\pi \rho
- \frac{1}{2} \nabla^2 \bar{\phi}  \; ,
\label{pares_eq_h00}\\
  \nabla^2 \bar{\phi} - m_{SF}^2 \bar{\phi} &=& - 8\pi \alpha\rho \; ,
\label{pares_eq_phibar}
\end{eqnarray}
we have set $\langle\phi\rangle=(1+\alpha)/G_N$ 
and $\alpha \equiv 1 / (3 + 2\omega)$.  
We 
are considering that the
influence of dark matter is due to a boson field of mass $m_{SF}$ governed by
Eq.\ (\ref{pares_eq_phibar}), that is the modified Helmholtz equation.
Equations (\ref{pares_eq_h00}) and (\ref{pares_eq_phibar}) represent
the Newtonian limit of a STT with arbitrary potential $V(\phi)$ and function
$\omega(\phi)$ that where Taylor expanded around $\langle\phi\rangle$.
The resulting equations are then distinguished by the constants
$G_N$ (the local gravitational constant), 
$\alpha$, and $\lambda=h_P/m_{SF}c$. Here $h_P$ is Planck's constant.

Note that Eq. (\ref{pares_eq_h00}) can be cast as a Poisson equation for
$\psi \equiv (1/2) ( h_{00} + \bar{\phi}/ \langle \phi\rangle)$, 
\begin{equation}
\nabla^2 \psi = 4\pi \frac{G_N}{1+\alpha} \rho \; . \label{pares_eq_psi}
\end{equation}

The next step is to find solutions for this new Newtonian potential given 
a density profile, that is, to find the so--called potential--density pairs. 
General solutions to Eqs. (\ref{pares_eq_phibar}) and (\ref{pares_eq_psi})
can be found in terms of the corresponding Green functions,
and the new Newtonian potential is (see \cite{mar2004,mar2005} for details)
\begin{eqnarray}
\Phi_N  \equiv \frac{1}{2} h_{00}
&=& - \frac{G_N}{1+\alpha} \int d{\bf r}_s
\frac{\rho({\bf r}_s)}{|{\bf r}-{\bf r}_s|} \nonumber \\
&& -\alpha \frac{G_N}{1+\alpha} \int d{\bf r}_s \frac{\rho({\bf r}_s)
{\rm e}^{- |{\bf r}-{\bf r}_s|/\lambda}}
{| {\bf r}-{\bf r}_s|} + \mbox{B.C.} \label{pares_eq_gralPsiN}
\end{eqnarray}
The first term of Eq. (\ref{pares_eq_gralPsiN}), given by $\psi$, is the
contribution of the usual Newtonian gravitation (without scalar
fields), while information about the scalar field is contained in the
second term, that is, arising from the influence function determined by the
modified Helmholtz Green function, where the coupling $\omega(\alpha$) enters
as part of a source factor.

\subsection{Cosmological evolution equations using a static STT}
To simulate cosmological systems,  the expansion of the universe has to be
taken into account.
Also, to determine the nature of the cosmological model we need to determine
 the composition of the
universe, i. e., we need to give the values of $\Omega_i$ for each component $i$, 
taking into account
in this way all forms of energy densities that exist at present.

If a particular kind of energy density is described by an equation of state of the form
$p=w \rho$, where $p$ is the pressure and $w$ is a constant, then the equation for energy
conservation in an expanding background, $d(\rho a^3)=-pd(a^3)$, can be integrated to
give $\rho \propto a^{-3(1+w)}$. Then, the Friedmann equation for the expansion factor $a(t)$
is written as
\begin{equation}
H^2 \equiv \frac{\dot{a}^2}{a^2} = H_0^2 \sum_i \Omega_i \left(\frac{a_0}{a}\right)^{3(1+w_i)} - \frac{k}{a^2}
\end{equation}
where $k$ characterizes the geometry of the universe ($k=0$ for a flat universe), and
$w_i$ characterizes the equation of state of specie $i$. Here, $\Omega_i \equiv \rho_i/\rho_c$,
with $\rho_c = 3H^2/8\pi G_N$. The most familiar forms of energy
densities are those due to pressureless matter with $w_i=0$ (that is, nonrelativistic matter
with rest-mass-energy density $\rho c^2$ dominating over the kinetic-energy density
$\rho v^2/2$) with $\Omega_{DM}\approx 0.22$ and radiation with $w_i=1/3$ and 
$\Omega_R \approx 2\times 10^{-5}$.  
The density parameter contributed today
by visible, nonrelativistic, baryonic matter in the universe is $\Omega_B \approx 0.04$.
There is also another main component, the energy density associated to
a cosmological constant, $\Omega_\Lambda \approx 0.74$ with an equation of state such
that $w_i = -1$.
In this work we will consider a model with only one energy density contribution.
One which is a pressureless and
nonbaryonic dark matter  with contribution given by $\Omega_{DM}$ 
that does not couple with radiation.
The above equation for $a(t)$ in this case becomes
\begin{equation}
\frac{\dot{a}^2}{a^2} = H_0^2 
\Omega_{DM} \left(\frac{a_0}{a}\right)^{3} 
- \frac{k}{a^2}
\end{equation}

Here, we employ a cosmological model with a static scalar field which is consistent with the 
Newtonian limit given by Eq. (\ref{pares_eq_gralPsiN}).
Thus, the scale factor, $a(t)$,  is given by the following Friedman model (see a more general
case in \cite{mar2008}),
\begin{equation} \label{new_friedman}
a^3 H^2= H_{0}^{2} \left[\frac{\Omega_{DM0} }{1+\alpha} 
+  \left(1-\frac{\Omega_{DM 0} 
}{1+\alpha} 
\right) \, a  \right]
\end{equation}
where 
$\Omega_{DM 0}$ 
is the dark matter density evaluated at present, respectively.   
We notice that the source of the cosmic evolution is deviated by the term 
$1+\alpha$ when compared to the standard Friedman-Lemaitre 
model. Therefore, it is convenient to define a new dark matter density parameter by 
$\Omega_{DM}^{(\alpha)} \equiv \Omega_{DM}/(1+\alpha)$. 
This new density parameter is such that always
$\Omega_{DM}^{(\alpha)}  =1$,
which implies a flat dark matter dominated universe, and this shall be assumed 
in our following computations.
For positive values 
of $\alpha$, a flat cosmological model demands to have a factor $(1+\alpha)$ more energetic 
content ($\Omega_{DM}$)  
than in standard FCDM cosmology. 
On the other hand, for negative values of  
$\alpha$ one needs a factor $(1+\alpha)$  less $\Omega_{DM}$ 
to have a flat universe.  To be consistent 
with the CMB  spectrum and structure formation numerical 
experiments, cosmological constraints must be applied on $\alpha$ in order for it to 
be within the range $(-1,1)$ \cite{Nagata2002,Nagata2003,Shirata2005,Umezu2005}.  

In the Newtonian limit of STT of gravity, 
the Newtonian motion equation  for a particle $i$ is written as
\begin{equation} \label{eq_motion}
\ddot{\mathbf{x}}_i + 2\, H \, \mathbf{x}_i = 
-\frac{1}{a^3} \frac{G_N}{1+\alpha} \sum_{j\ne i} \frac{m_j (\mathbf{x}_i-\mathbf{x}_j)}
{|\mathbf{x}_i-\mathbf{x}_j|^3} \; F_{SF}(|\mathbf{x}_i-\mathbf{x}_j|,\alpha,\lambda)
\end{equation}
where $\mathbf{x}$ is the comovil coordinate, and  the sum includes all  
periodic images of particle $j$,  and $F_{SF}(r,\alpha,\lambda)$ is
\begin{equation}
F_{SF}(r,\alpha,\lambda) = 1+\alpha \, \left( 1+\frac{r}{\lambda} \right)\, e^{-r/\lambda}
\end{equation}
which,  for small distances compared to $\lambda$,  is 
$F_{SF}(r<\lambda,\alpha,\lambda) \approx 1+\alpha \, \left( 1+\frac{r}{\lambda} \right)$ and, 
for long 
distances, is  $F_{SF}(r>\lambda,\alpha,\lambda) \approx 1$, as in Newtonian physics. 

We now analyze the general effect that the constant $\alpha$ has on the dynamics.  
The role of $\alpha$ in our approach is
as follows.   On one hand, to construct a flat model  
we have set the condition $\Omega_{DM}^{(\alpha)}  =1$, which 
implies  having $(1+\alpha)$ times the energetic content of the standard FCDM model.
 This essentially means that
we have an increment by a factor of  $(1+\alpha)$ times the amount of matter, 
for positive values of  $\alpha$, or a 
reduction of the same factor for negative values of  $\alpha$. Increasing or reducing 
this amount of matter affects 
the matter term on the  r.h.s. of the equation of 
motion (\ref{eq_motion}), but the amount affected cancels out with the term $(1+\alpha)$ in the 
denominator of  (\ref{eq_motion}) stemming from the new Newtonian potential. 
On the other hand, the factor $F_{SF}$ augments (diminishes) for positive (negative)  
values of $\alpha$ for small distances compared to  $\lambda$, resulting in more (less) structure formation for positive (negative) values of $\alpha$ compared to the FCDM model.  
For $r\gg \lambda$ the dynamics is essentially Newtonian.

\section{Results}
\subsection{Cosmological simulations of the Santa Barbara cluster}
In this section, we present results of the cosmological simulations of a FCDM universe
with and without SF contribution. The initial condition of the system corresponds to the well known
{\em Santa Barbara cluster} data that we get from the 
Heitmann's Cosmic Data Bank web page (\url{http://t8web.lanl.gov/people/heitmann/test3.html}). 
The initial condition uses a box 64 Mpc  size, $128^3$ particles and start at redshift $z=63$.

In 1999 Frenk et al. \cite{Frenk1999} reported results of an extensive code comparison project involving
twelve different codes.
The aim of the project was to compare different techninques for simulating the formation of a cluster
of galaxies (by now widely known as the {\em Santa Barbara cluster}) in a flat cold dark matter universe
and to decide if the results from different codes were consistent and reproducible. We have repeated this
test but restricted ourselves only to the dark matter component of the test. 
The cosmological parameters
used in the simulation are as follows. We used 128$^3$ particles with initial positions and velocities with
$\Omega_{DM}=1$, $\Omega_\Lambda=0$, $\Omega_B=0$, 
$H_0=50$ km/s/Mpc, box size = 32 Mpc$/h$,
$\sigma_8=0.9$ for the present-day linear rms mass fluctuation in spherical top spheres of radius 16 
Mpc. The FCDM universe contents is given in units of the critical density, $\rho_c$.
The initial fluctuation spectrum was taken to have an asymptotic spectral index, $n=1$, and the shape
parameter $\Gamma=0.25$, the value suggested by observations of large-scale structure \cite{Efstathiou1992}.
See \cite{Frenk1999} and Heitmann's web page above for full details.
In the shown snaps at $z=0$, Fig. 1, the big cluster near the center of the frame is the SB cluster.

The
particle masses are $\approx 0.434\times 10^{10}$ M$_\odot/h$. 
The individual softening length
was 50 kpc$/h$. These choices of softening length are consistent
with the mass resolution set by the number of particles.

We now present results for the FCDM model previously described. 
Because the visible component is the smaller one and given our interest to
test the consequences of including a SF contribution to the evolution equations,
our model excludes gas particles, but all its mass has been added to the dark matter. 
We restrict the values of $\alpha$ to the interval $(-1,1)$ 
  \cite{Nagata2002,Nagata2003,Shirata2005,Umezu2005}  and  use $\lambda=1$ Mpc$/h$, since 
this scale turns out to be an intermediate scale between the size of the clump groups and 
the separation of  the formed groups.

\begin{figure}
\begin{minipage}{6in}
\begin{center}
\includegraphics[width=2.75in]{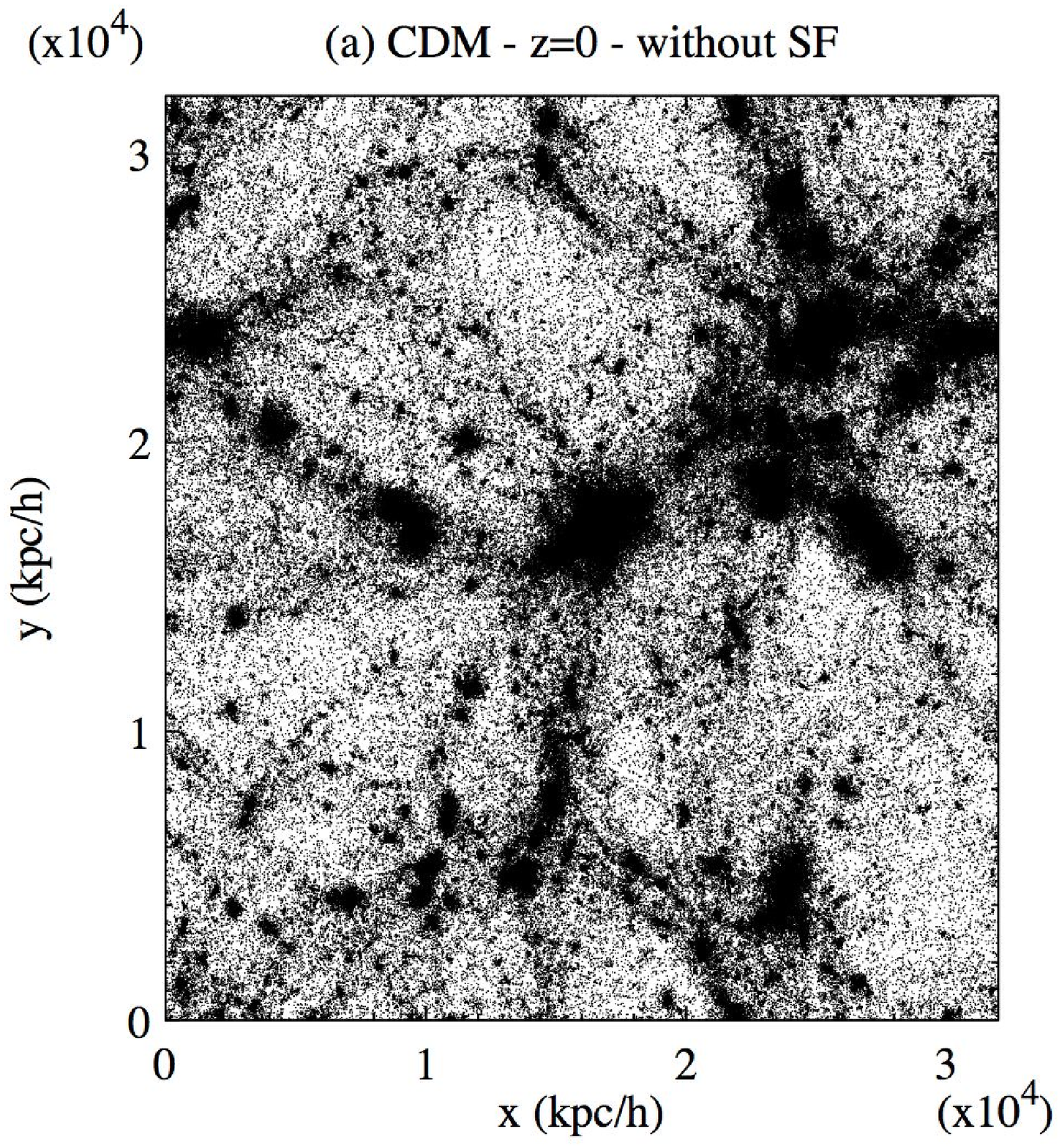}
\includegraphics[width=2.75in]{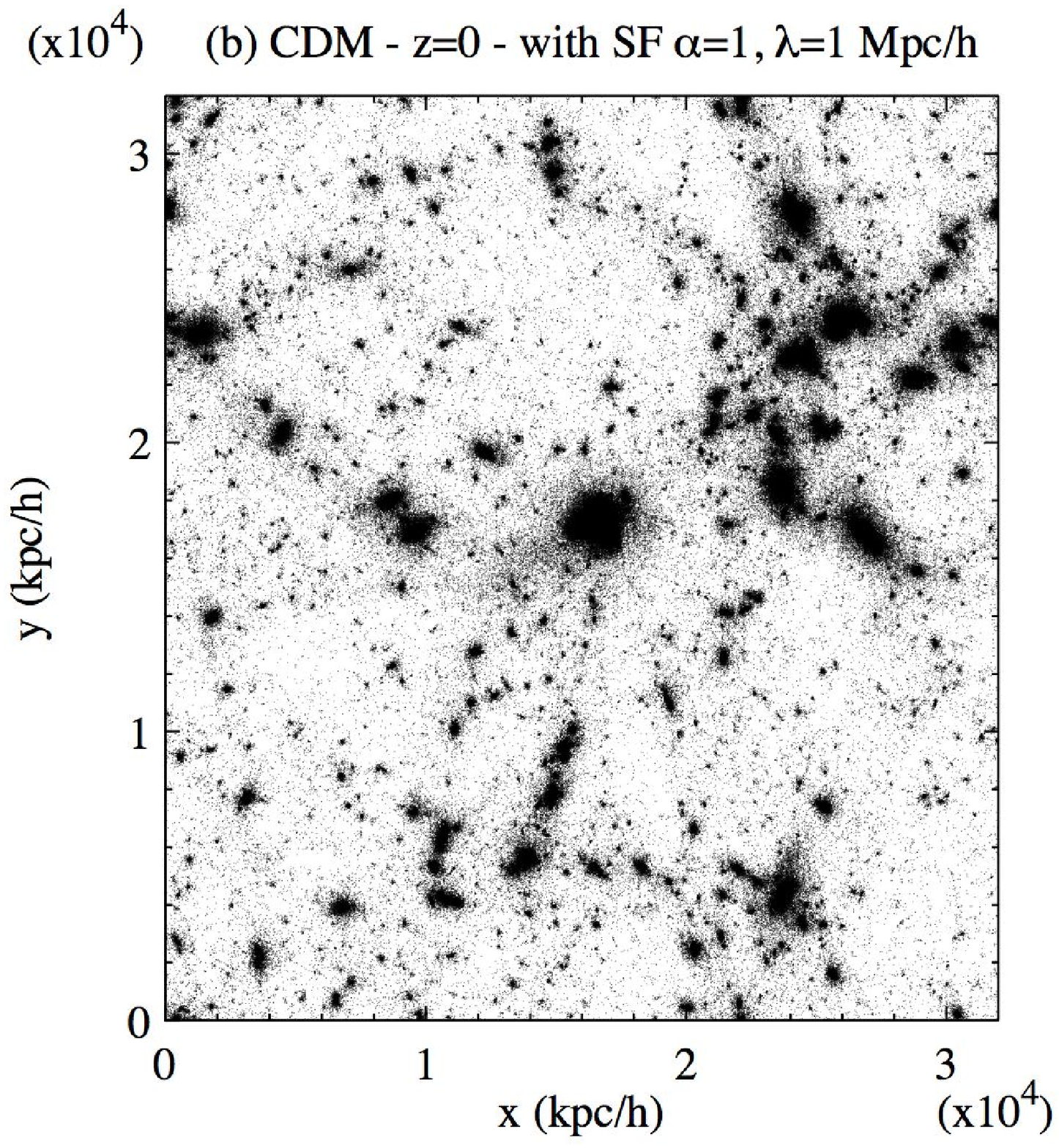}
\includegraphics[width=2.75in]{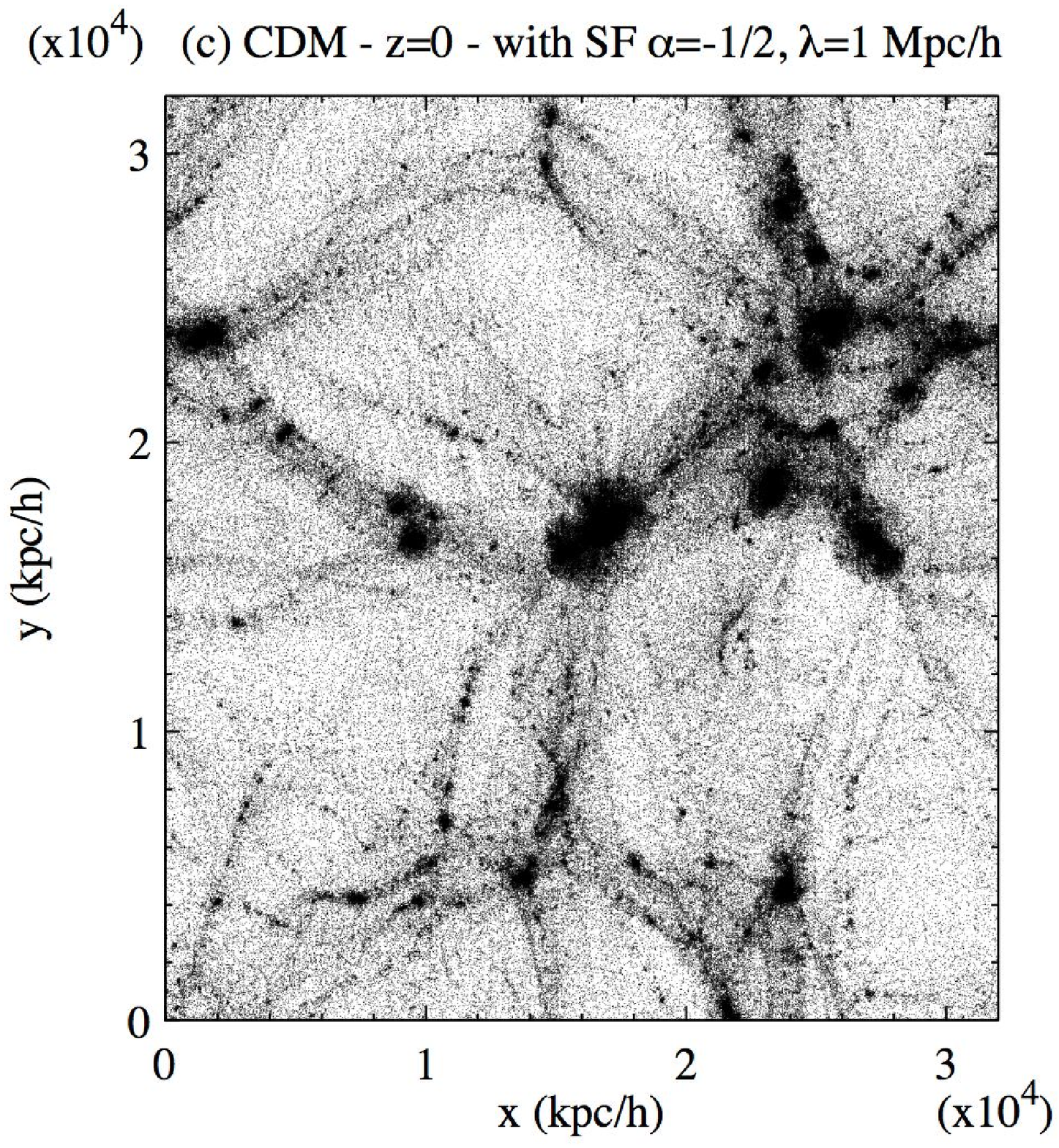}
\includegraphics[width=2.75in]{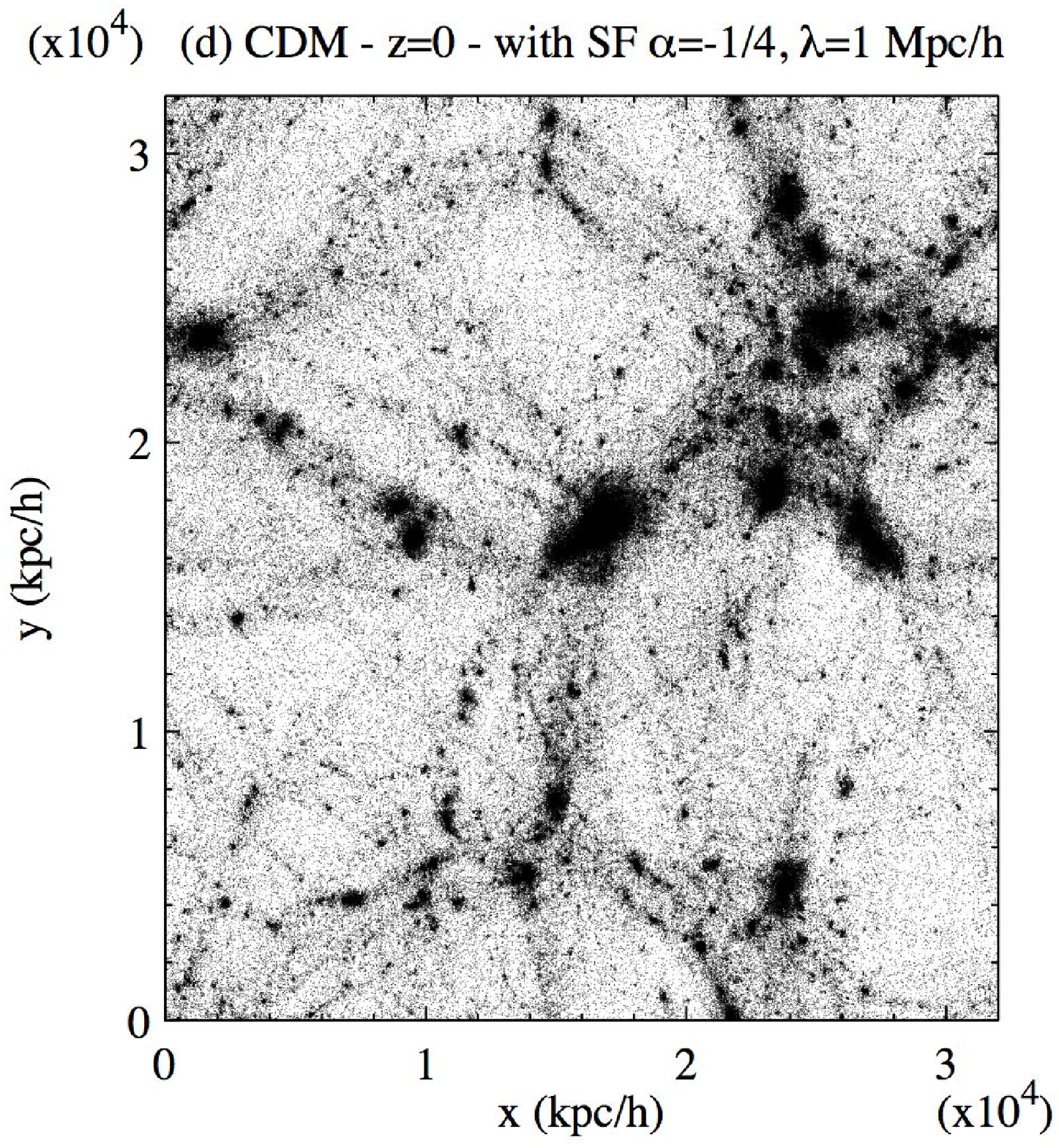}
\end{center}
\end{minipage}
\caption{$x$--$y$ snapshots at $z=0$ of a FCDM universe.  See text for details. }
\end{figure}
In Fig. 1 we show $x$--$y$ snapshots at redshift $z=0$ of our  FCDM model. 
Fig. 1 (a) presents the standard case without SF, i.e., the interaction between bodies  is through
the standard Newtonian potential.
In (b) we show the case with $\alpha=1$, $\lambda=1$ Mpc$/h$.
In (c)  $\alpha=-1/2$, $\lambda=1$ Mpc$/h$.
In (d) $\alpha=-1/4$, $\lambda=1$ Mpc$/h$.  
One notes clearly how the SF modifies the matter  structure of the system. The most
dramatic cases are (b) and (c) where we have used 
$\alpha=1$ and $\alpha=-1/2$, respectively. 
Given the argument  at the end of last section, in the case of (b), for $r \ll \lambda$,
the effective gravitational pull has been  augmented by a factor of $2$, 
in contrast to case (c) where it has diminished  by a factor of 1/2; in model (d) the pull 
diminishes only by a factor of 3/4. That is why one observes for $r < \lambda$ more structure 
formation in (b), less in (d), and lesser in  model (c).  
The effect is  then, for a growing positive $\alpha$, 
to speed up  the growth of perturbations, then of halos and then of clusters, whereas negative 
$\alpha$ values ($\alpha \rightarrow -1$) tend to slow down the growth. 

\begin{figure}
\includegraphics[width=3.1in]{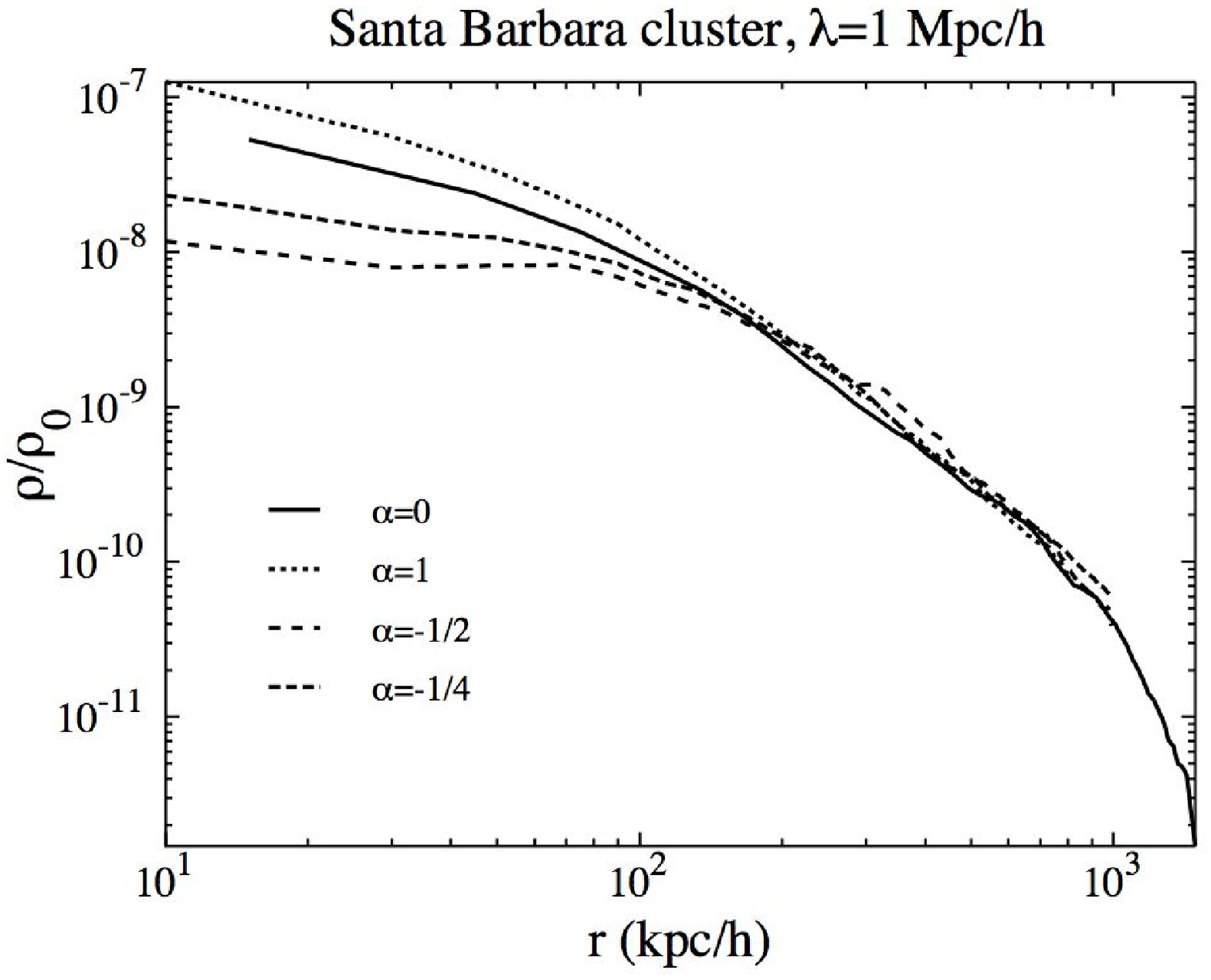}
\includegraphics[width=3.1in]{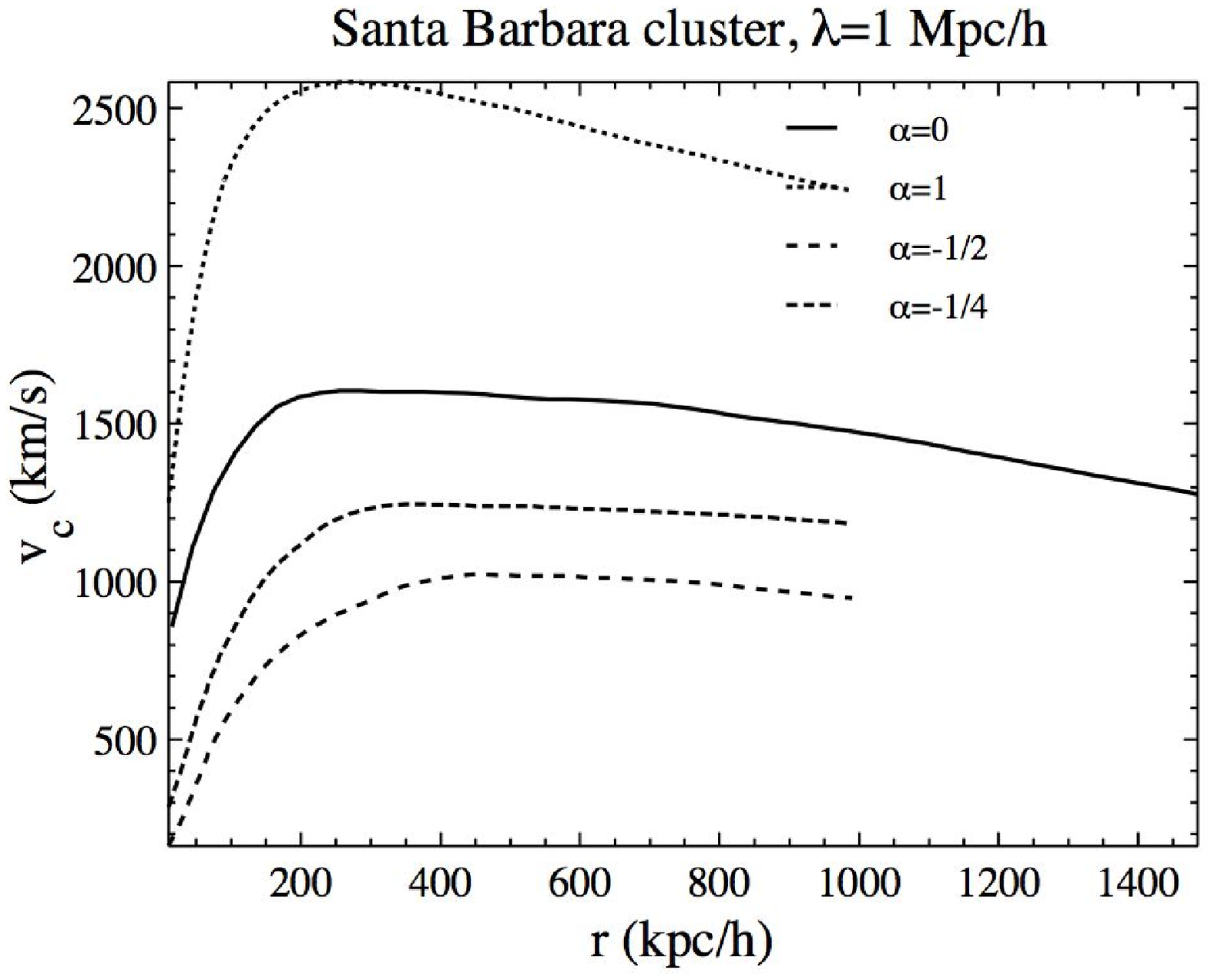}
\caption{(a) Density profiles of the SB cluster
at $z=0$. 
 The cluster is located at the center in Fig 1. 
Vertical scale is in units of $\rho_0=10^{10} \mbox{M}_\odot h^{-1} / (h^{-1}\mbox{kpc})^3$.
(b) The corresponding circular velocity.}
\end{figure}
Next, we find the groups in the system using a friend-of-friend (FOF) algorithm
and select one of the most massive ones. The chosen group is located
approximately at the center in Fig. 1, the SB cluster.
The group was analyzed by obtaining their density profiles (Fig. 2(a))
and circular velocities (Fig. 2(b)).
The  more cuspy case is for $\alpha = 1$ and
the less cuspy is for $\alpha=-1/2$. 
The circular velocity curves where computed using 
$v_c^2=G_N M(r)/r$. The case with $\alpha=1$ corresponds to higher values of $v_c$, 
since this depends on how much accumulated mass there is at a distance $r$ and this
is enhanced by the factor 
$F_{SF}$ for positive values of $\alpha$.


\subsection{Equation of state of a dark matter halo}
We finish this work by discussing a possible way to find a dark matter halo equation of state (EOS).
The EOS may be useful to characterize more completely the state of a dark matter halo and could
be a way to discriminate dark matter models.
 
In general relativity both density and pressure contribute to modify the space-time geometry.
For a static and spherically symmetric system in equilibrium 
like a dark matter halo we can not {\em a priori}
neglect the pressure contribution. Therefore, and in the Newtonian limit we must solve the equation 
(see Misner et al. \cite{Misner1973})
\begin{equation} \label{NewtonianPoisson_eq}
\frac{1}{r^2}\frac{d}{dr}\left(r^2\frac{d\Phi}{dr}\right)  \approx 4 \pi G (\rho  + p)
\end{equation}
together with the equation of
hydrostatic equilibrium condition,
\begin{equation}\label{HydrostaticEquilibrium_eq}
\frac{dp}{dr} = - (\rho+p) \frac{d\Phi}{dr}
\end{equation}
where $p(r)=p_r(r) + 2p_t(r)$ and we have assumed spherical symmetry.
We also use the flatness condition on the circular velocity at large distances,
\begin{equation}
V_{c0}^2=r \frac{d \Phi_N}{dr} = \mbox{\ \ constant}
\end{equation}
to construct the boundary condition
\begin{equation}
\left. \frac{dp}{dr}\right|_{r=R_H} = -(\rho(R_H)+p(R_H))  \frac{V_{c0}^2}{R_H}
\end{equation}
where $R_H$ is the size of the system. The other condition is
\begin{equation}
 p(R_H) =0
\end{equation}

\begin{figure}\label{eos_nfw_piso}
\includegraphics[width=3.1in]{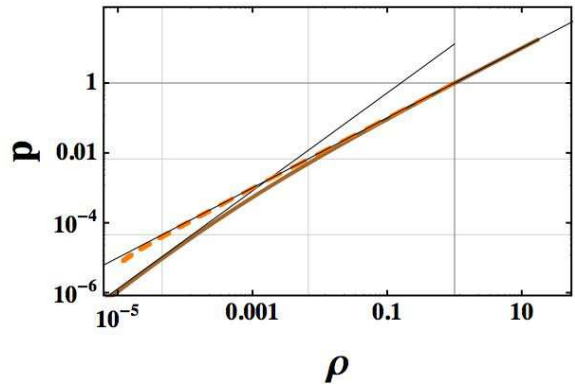}
\caption{Equation of state for two density profiles. Pseudo-isothermal profile which
has a power law behavior, $p \sim \rho^n$ with $n=1$ 
and the NFW profile which has two power law behavior,
one with $n=1.4$ and another with $n=1$. Pressure and density are in geometrical
units}
\end{figure}

Results for two density profiles are given in figure \ref{eos_nfw_piso}.
The dashed line is for the isothermal density profile given by,
\begin{equation}\label{rho_iso_eq}
\rho_{ISO} = \frac{\rho_c}{1+(r/r_c)^2}
\end{equation}
where $\rho_c$ is the core density and $r_c$ is the scale length of the matter distribution given by
the isothermal profile. Whereas solid line is the Navarro-Frenk-White (NFW) density profile given by
\begin{equation}\label{rho_nfw_eq}
\rho_{NFW} = \frac{\rho_{cs}}{(r/r_s)(1+r/r_s)^2}
\end{equation}
where $\rho_{cs}$ is the NFW typical density and $r_s$ is the scale length of the matter distribution given by the NFW profile.

\begin{figure}\label{santabarbara}
\includegraphics[width=3.1in]{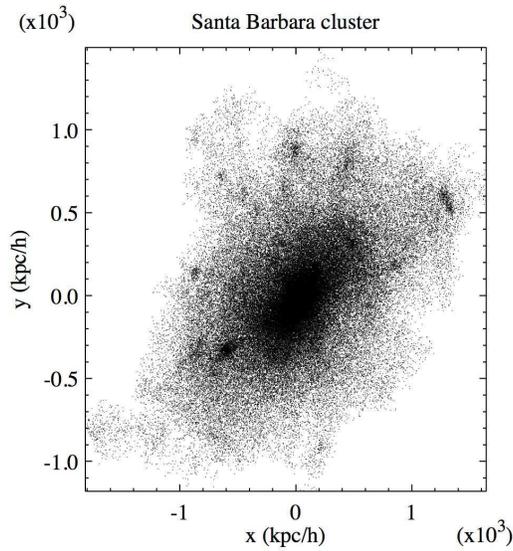}
\caption{The SB cluster extracted from the numerical simulation of the CDM
model using a FOF algorithm.}
\end{figure}

\begin{figure}\label{eos_santabarbara}
\includegraphics[width=3.1in]{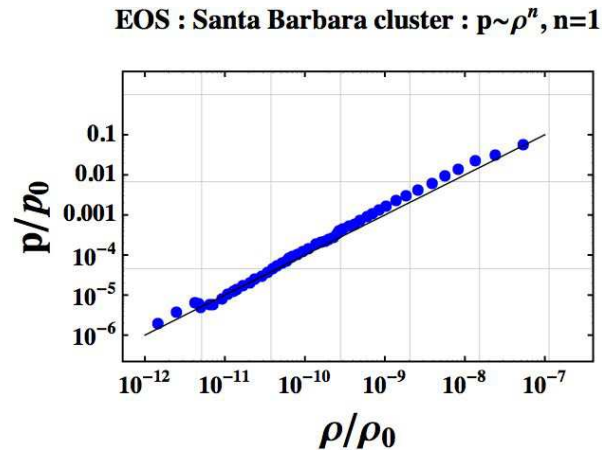}
\caption{Equation of state for SB cluster (dots). $p_0$ is the unit of pressure 
$6.77\times 10^{-13}$ Pascals. The solid line is for an equation of state $p\sim \rho^n$ with $n=1$.}
\end{figure}

We may observe that NFW EOS has two power laws, i.e., behavior $p\sim \rho^n$, for low densities
$n=1.4$ and for high densities $n=1$ that is the behavior of the isothermal EOS. We may compare
this results with the EOS of the SB cluster, shown in Fig. \ref{santabarbara}. Where
we have assumed in order to do the calculations, that $p=\rho \sigma_r^2$, here $\sigma_r^2$ is
the radial dispersion of velocities of the cluster.
This numerical EOS is shown in figure \ref{eos_santabarbara}. We may notice the power law behavior,
$n=1$ which corresponds to an isothermal density profile or to the high density case for the 
NFW EOS case.


\section{Conclusions}
The theoretical
scheme we have used 
is compatible with local observations because we have defined
the background field constant 
$<\phi>  =  G_{N}^{-1} (1+\alpha)$.  A direct consequence of  
the approach is that  the amount of matter (energy) has to be increased 
for positive values of $\alpha$ and diminished  for negative values of $\alpha$ 
with respect to the standard FCDM model 
in order to have a flat cosmological model. Quantitatively, our model demands to 
have $\Omega/ (1+\alpha) =1$ and this changes the amount of dark matter and 
energy of the model for a flat cosmological model, as assumed.   
The general gravitational effect is that  the interaction including  the SF changes by a factor 
$F_{SF}(r,\alpha,\lambda) \approx 1+\alpha \, \left( 1+\frac{r}{\lambda} \right)$ for $r<\lambda$ in 
comparison with the Newtonian case. Thus, for $\alpha >0$ the growth of structures speeds up  
in comparison with the Newtonian case.  For the   $\alpha <0$ case the effect is to diminish 
the formation of structures.  For $r> \lambda$ the dynamics is essentially Newtonian.

Additionally we have found numerically and EOS for the SB cluster. However, we assume
that $p = \rho \sigma_r^2$, where $\rho$ and $\sigma_r$ were obtain from SB cluster particle
data. We leave for a future paper the numerical computation of $p_r$ and $p_t$ from the
pressure tensor of the SB cluster and more detailed analysis. 
We have compared the power-law behaviour ($p\sim \rho^n$) of the 
SB cluster EOS with the corresponding behavior of an EOS obtained
solving Eqs. (\ref{NewtonianPoisson_eq}) and (\ref{HydrostaticEquilibrium_eq}) for two density
profiles, isothermal profile (\ref{rho_iso_eq}) and NFW profile (\ref{rho_nfw_eq}).  We see that
the SB cluster is like the isothermal profile for some range of lower densities. 
Even more NFW profile has two power-law, for
low densities behaves as $n=1.4$ and for high densities as $n=1$ that corresponds two
the isothermal profile.


\bigskip
{\it Acknowledgements: }
This work was supported by CONACYT, grant number I0101/131/07 C-234/07, IAC collaboration. 
The simulations were performed in the UNAM HP cluster {\it Kan-Balam}.

\end{document}